\documentclass[a4paper,11pt]{article}
\usepackage{pos}
\usepackage{amsmath}
\title{A study of bent jets in active galactic nuclei at parsec scales}
\ShortTitle{Bent jets in AGN}

\author*[a, b]{V. A. Makeev}
\author[c, b, a]{Y. Y. Kovalev}
\author[d, b]{A. B. Pushkarev}

\affiliation[a]{Moscow Institute of Physics and Technology,
  Institutsky per. 9, Dolgoprudny 141700, Russia}

\affiliation[b]{Lebedev Physical Institute,
Leninsky prospekt 53, 119991 Moscow, Russia}
\affiliation[c]{Max-Planck-Institut für Radioastronomie,
Auf dem Hügel 69, D-53121 Bonn, Germany}

\affiliation[d]{Crimean Astrophysical Observatory, 
Nauchny 298688, Crimea, Russia}

\emailAdd{makeev.va.ph@gmail.com}

\abstract{Very long baseline interferometry (VLBI) observations show that some active galactic nuclei (AGN) jets exhibit bending even at parsec scales. The nature of bending is comprehensively analysed only for a small number of individual AGN, and the overall  trends in shape of the substantially curved jets are unclear. In this work, we analyse outflows in AGN on the basis of publicly available multi-frequency VLBI images. Nearly 73~000 images of about 11~000 AGN are studied. Our research reveals that about 5\% of them show a significantly curved jet structure. We characterize the jets geometry by fitting total intensity ridge lines constructed at all available frequencies and epochs with a set of simple models and suggest possible scenarios explaining the observed bending.}

\FullConference{%
  15th European VLBI Network Mini-Symposium and Users' Meeting (EVN2022)\\
  11-15 July 2022\\
  University College Cork, Ireland
}

\begin{document}
\maketitle

\section{Introduction}
    Decades of observations of active galactic nuclei (AGN) jets with Very Long Baseline Interferometry (VLBI) reveal compellingly that some of them exhibit bending \cite{sbj1, sbj2, sbj3} which is present at the entire range of the observed scales, from parsecs to kilo-parsecs \cite{kpc, pckpc}. Different physical mechanisms can cause this phenomenon, visually enhanced by projection effects. The possibilities include pure precession of the jet nozzle \cite{precbh, precad}, Kelvin-Helmholtz instability \cite{KH}, interaction with ambient medium \cite{ismcol} or their combinations. 
    
\begin{figure}[h]
        \centering
        \includegraphics[width=1.\textwidth]{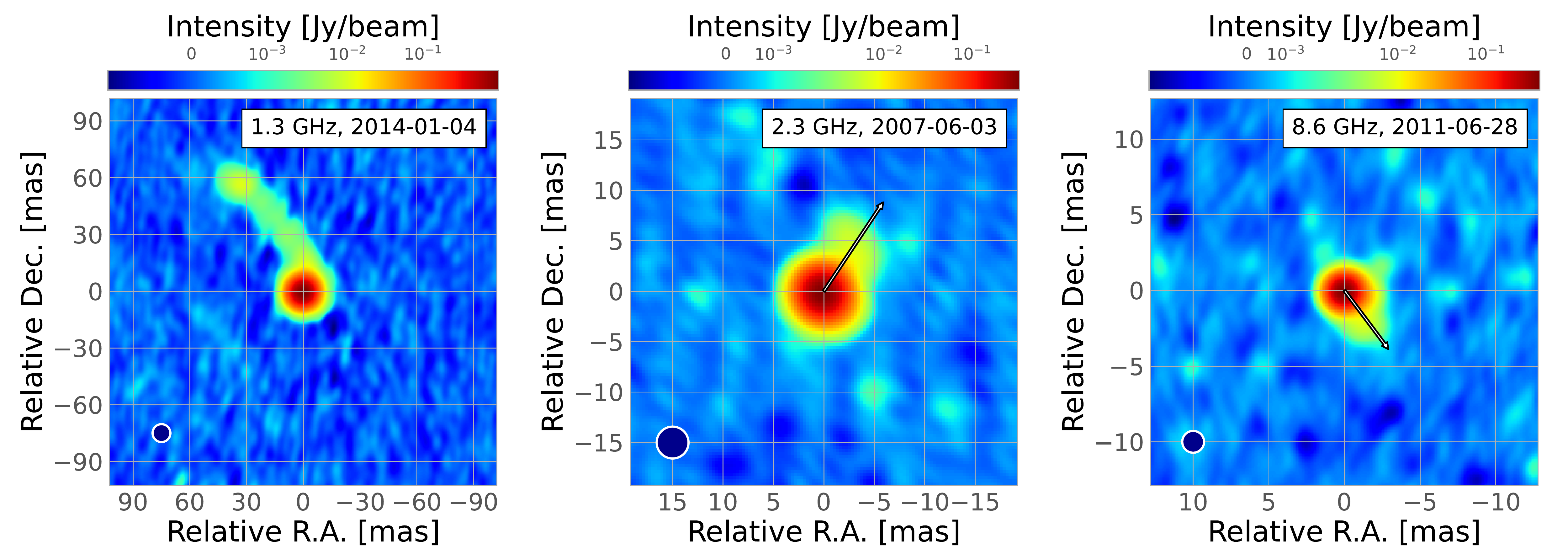}
        \caption{The quasar J1327+2210 (TXS~1324+224): an example of the source with significantly different jet directions probed at different scales by VLBI observations at 1.4, 2.3 and 8.6~GHz. This reflects a helix-shaped structure of the entire jet. The black arrows (center and right panels) indicate the corresponding jet position angle at a given frequency.
        The restoring beam at the half-power level is shown in the bottom left corners.}
        \label{im:padif}
    \end{figure}

    It is crucial to analyse the jet morphology probed by observations performed at different frequencies. The observations taken at longer wavelengths may not reveal smaller-scale structures which are seen at shorter wavelengths: at individual frequencies, the jet may be interpreted as straight, but multi-frequency analysis may establish a strong difference in position angles between these frequencies (Fig.~\ref{im:padif}), which is an evidence of the jet bending. 
    
    Our analysis aims at finding AGN with significantly bent jets at parsec scales using publicly-available multi-frequency VLBI data. Additionally, we describe the observed jet morphologies with simple geometrical models for further numerical interpretation and association with physical models.

\section{Observational data}

    In this paper, we utilise the Astrogeo\footnote{\href{http://astrogeo.org/vlbi_images/}{http://astrogeo.org/vlbi\_images/}} VLBI FITS image database which contains more than 100~000 brightness maps of about 17~000 compact radio sources collected from several major VLBI surveys. We investigate about 73~000 images of approximately 11~000 AGN in total. On average, each object has been observed at eight epochs and two frequencies. The majority of images in the database do not reveal a resolved extended  jet structure. To analyse the bending patterns, we perform a simple filtering procedure to avoid that class of images. The filtering process is described in Section 3.

\section{Images filtering}

    To select images with the extended jet structure, we utilised the following algorithm:
    
    \begin{enumerate}
        \item The image is fitted with a 2-dimensional Gaussian model.
        \item The model is subtracted from the image.
        \item The remaining flux density is calculated.
        \item If the remaining flux is greater than the manually set optimal value ($5.5 \sigma_{\text{noise}} $), the target object is considered to have an extended structure.
    \end{enumerate}
    
    The filtering procedure reduced the initial data set by 52\%. The remaining part contains about 35~000 images of 5500 sources with resolved jet structure.

\section{Extracting AGN with bent jets}

    To describe the geometry of AGN jets, we first compute their ridge lines \cite{rline} at all available epochs and frequencies. To find candidates which potentially have a significantly bent jet structure, we stack the obtained ridge lines and model them with the linear function $\phi = a r + b$, where $r$ and $\phi$ are the distance from the core and the position angle, respectively. The objects having the total change of the position angle in linear approximation, $\Delta \phi = a (r_{\text{max}} - r_{\text{min}}) > 20 ^{\circ}$ are considered as significantly bent. In order to avoid uncertainty of the jet direction in the core region, $r_{\text{min}} > 0$ is determined as the minimal radius of the concentric circle covering at least one map pixel with intensity less than $2 \sigma_{\text{noise}}$. To find $r_{\text{max}}$, we calculate mean intensities in the 5\%-map-pieces centered in each ridge line point. The largest radial distance at which the mean intensity is greater than $2 \sigma_{\text{noise}}$ defines $r_{\text{max}}$. The angle $\Delta\phi$ can only be used to roughly approximate the apparent angle of bending because the morphologies of the curved jets are often substantially different from those described by a linear dependence. 

    The search procedure described above has resulted in identifying about 1000 candidates. Subsequent manual selection revealed that 216 of them have significantly bent jets.  

\section{Jet structure patterns}

\begin{figure}[h]
        \centering
        \includegraphics[width=0.70\textwidth]{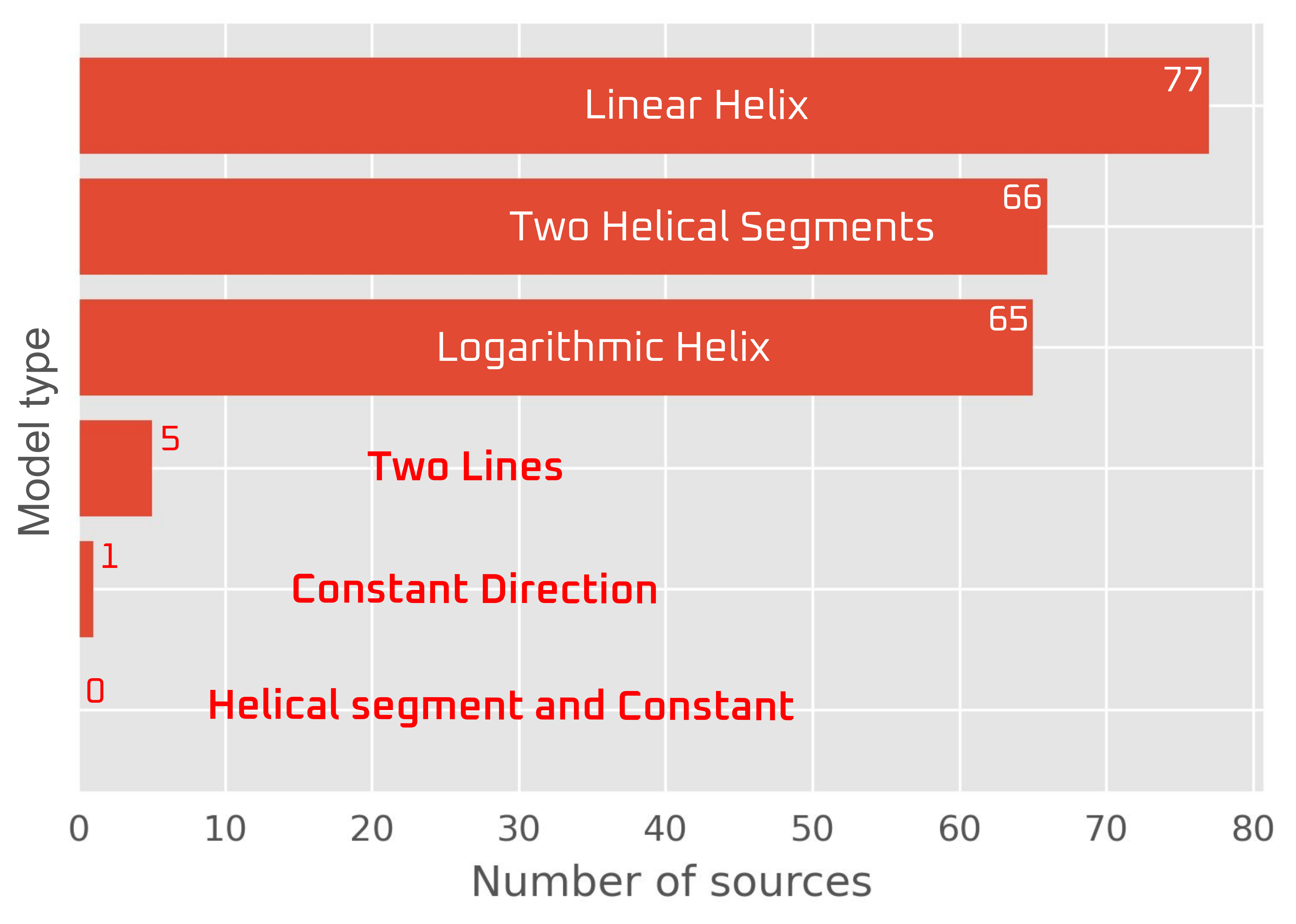}
        \caption{Distribution of best-fit models of AGN jets.}
        \label{im:mdist}
    \end{figure}

    To study geometry of jets, we fit stacked ridge lines with the following simple models: line of the constant position angle (Eq. \ref{mod1}), linear helix (Eq. \ref{mod2}), logarithmic helix (Eq. \ref{mod3}), combination of helical segment and constant position angle segment (Eq. \ref{mod4}), combination of two linear helical segments (Eq. \ref{mod5}), combination of two consecutive linear segments (Eq. \ref{mod5} with substitutions \hbox{$\phi \mapsto y = r \sin{\phi}$} and \hbox{$r \mapsto x = r \cos{\phi}$} ).

    \begin{equation} \label{mod1}
    \phi (r) = \phi_0,
    \end{equation}

    \begin{equation}\label{mod2}
        \phi (r) = a + b  r,
    \end{equation}
    
    \begin{equation}\label{mod3}
        \phi (r) = a \ln{\frac{r}{b}},
    \end{equation}
    
    \begin{equation}\label{mod4}
        \phi (r) = 
        \begin{cases}
            a + b  r & r \leq r_0 \\
            \phi_0 & r > r_0
        \end{cases},
    \end{equation}

    \begin{equation} \label{mod5}
        \phi (r) = 
        \begin{cases}
            a + b  r & r \leq r_0 \\
            a + b r_0 + c (r - r_0) & r > r_0
        \end{cases},
    \end{equation}
where $a$, $b$, $c$, $r_0$ and $\phi_0$ are parameters optimized for each model. As in Section~4, we perform fitting in the polar coordinate system centered at the core position. Given the optimal parameters, we chose the best model based on the Bayesian Information Criterion. The final distribution of the best-fit models is presented in Fig.~\ref{im:mdist}.

\section{Discussion and future work}

    Fig.~\ref{im:mdist} shows that three models are dominant amongst the others: linear helix, logarithmic helix and a combination of two helical segments. If we consider the jet to be a 3-dimensional helix projected on the surface of the inclined cone with a constant opening angle, the linear helix model is more likely to be optimal when we look inside the cone: the position angle monotonically changes along the entire jet with approximately the same rate. The case of two linear helical segments is likely to represent the situation at which we look at the jet cone from outside. In this case the jet changes its positional angle rapidly only in the central region. Further down the flow the position angle change rate is reduced but it can still be described by a helix with other parameters. Logarithmic helix is likely to be the case of a non-helical jet.

    At the next step of our analysis, we plan to perform Monte-Carlo modeling of the jet morphology. This approach will help in interpreting the derived distribution (Fig.~\ref{im:mdist}) numerically and setting restrictions at geometrical and physical parameters of the initial helix. For example, such parameters are the cone inclination, the opening angle, the apparent jet velocity and the cone precession rate.

\section{Acknowledgements}

    We thank Astrogeo database contributors for making calibrated VLBI data publicly available and Leonid Petrov for maintaining the database. We are grateful to Elena Bazanova for language editing and Andrei Lobanov for valuable comments. This research was supported by the Russian Science Foundation \mbox{(project 21-12-00241)}.

\end{document}